\documentstyle[prl,aps,psfig,twocolumn]{revtex} 

\newcommand{\avg}[1]{\langle{#1}\rangle}

\begin{document}
\draft
\twocolumn[\hsize\textwidth\columnwidth\hsize\csname
@twocolumnfalse\endcsname
\title{Multiscale behaviour of volatility autocorrelations 
in a financial market}
\author{Michele Pasquini and Maurizio Serva}
\address{Istituto Nazionale Fisica della Materia \& 
Dipartimento di Matematica}
\address{Universit\`a dell'Aquila, I-67010 Coppito, L'Aquila, Italy}


\maketitle

\begin{abstract}
We perform a scaling analysis on NYSE daily returns.
We show that volatility correlations are power-laws
on a time range from one day to one year and,
more important, that they exhibit a multiscale behaviour.
\end{abstract}

\pacs{}

\narrowtext
]
It is well known that stock market returns are uncorrelated 
on lags larger than a single day, in agreement with the 
hypothesis of efficient market.
On the contrary, absolute returns have memory for longer times;
this phenomenon is known in financial literature as 
{\it clustering of volatility}.
In ARCH-GARCH models \cite{Engle,Jorion,AB}, volatility memory
is longer than a single time step but it decays exponentially 
while empirical evidence is for hyperbolic correlations
\cite{Taylor,DGE,BB,DLC,Baillie,Pagan}.
In this paper, we perform a scaling analysis of the standard deviation
of a new class of observables, 
the {\it generalized cumulative absolute returns}.  
This analysis clearly shows that volatility correlations 
are power-laws
on a time range from one day to one year and,
more important, that the exponent is not unique.
This kind of multiscale behaviour is known to be relevant 
in the theory of dynamical systems,
of fully developed turbulence and 
in the statistical mechanics of disordered 
systems (see \cite{PV} for a review) while it is a new concept
for financial modeling.

We consider the daily New York Stock Exchange (NYSE) index, 
from January 1966 to June 1998, for a total of $N=8180$ working days.
The quantity we consider is the (de-meaned) daily return, 
defined as \begin{equation}
r_t = \log{S_{t+1}\over{S_t}} - \avg{\log{S_{t+1}\over{S_t}}}
\end{equation}
where $S_t$ is the index value at time $t$ ranging from 1 to $N$, 
and $\avg{\cdot}$ is the average over the whole sequence.
The underlying daily volatility $\sigma_t$ is not directly observable, 
but it is indirectly defined by $r_t = \sigma_t \eta_t$. 
It is assumed that the $\eta_t$ are identically distributed
random variables with vanishing average and unitary variance.
The usual choice for the distribution of the $\eta_t$ 
is the normal Gaussian.
The observables directly related to the volatilities are the 
absolute returns $|r_t|$.

As pointed out by several authors \cite{Clark,Mandelbrot,MS1},
the distribution of returns is leptokurtic.
In \cite{Mandelbrot}, it was firstly proposed a
symmetric L\'evy stable distribution and more recently 
in \cite{MS1} it has been 
provided strong evidence for this fact. 
More precisely, in \cite{MS1} it is
shown that the distribution is L\'evy stable for high frequency returns
except for tails, which are approximately exponential.
The estimation is that the shape of a Gaussian 
is recovered only 
on longer scales, typically one month.

Our analysis is on low frequency data, and first of all we want to
verify that anomalous L\'evy scaling is not effective 
in a range of time from one day to one year.
We consider the {\it cumulative returns} $\phi_t(L)$, defined as
the sum of $L$ successive returns $r_t, \dots, r_{t+L-1}$, 
divided by $L$. Using NYSE data one can define
$N/L$ not overlapping variables of this type
and compute the standard deviation $\sigma(L)$.
The standard deviation is independently computed for $L$
 ranging from $1$ to $250$ (one year). 
Larger value of $L$ would imply insufficient statistics.
Assuming that $r_t$ are uncorrelated 
(or short range correlated),
it follows that $\sigma(L)$ has a power-law behaviour 
with exponent $0.5$ for large $L$, i.e. $\sigma(L)\sim L^{-\beta}$
 with $\beta=0.5$.
The exponent for the NYSE index turns out to be about $0.49$ 
(see fig. 1 and also see \cite{MS2}), 
according to the hypothesis of uncorrelated returns.
This value of the exponent also ensures 
that L\'evy scaling is not effective
in this range of time.

\begin{figure}
\mbox{\psfig{file=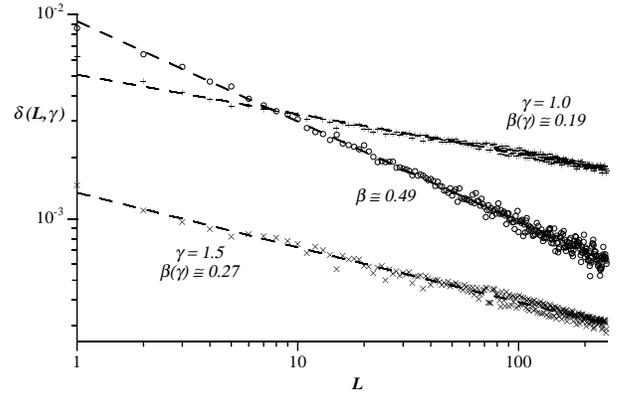,width=8cm,angle=270}}
\caption{Standard deviation $\delta(L,\gamma)$ of the generalized
cumulative absolute returns (2) as a function of $L$ on log-log scales
for $\gamma=1$ (crosses) and $\gamma=1.5$ (slanting crosses),
compared with the  standard deviation of 
the cumulative returns (circles).
The exponents of the best fit straight lines (dashed lines) are, 
respectively, 
$\beta(1) \simeq 0.19$, $\beta(1.5) \simeq 0.27$
and $\beta \simeq 0.49$.
}
\end{figure}

Let us introduce the {\it generalized cumulative absolute returns} 
defined as 
\begin{equation}
\chi_t(L,\gamma) = {1\over L} \sum_{i=0}^{L-1} 
\left|{r_{t+i}}\right|^\gamma
\end{equation}
where $\gamma$ is a real exponent and again,
these quantities are not overlapping. 
If the $\left|r_t\right|^\gamma$
are uncorrelated, one should find that 
the standard deviation $\delta(L,\gamma)$
has a power-law behaviour with exponent $0.5$.

On the contrary, a power-law autocorrelation function
with exponent $\alpha(\gamma)\le 1$
$\avg{\left|r_t\right|^\gamma \left|r_{t+L}\right|^\gamma}
-\avg{\left|r_t\right|^\gamma} \avg{\left|r_{t+L}\right|^\gamma}
\sim L^{-\alpha(\gamma)}$,
would imply that $\delta(L,\gamma)$ is
 a power-law with exponent $\beta(\gamma)=\alpha(\gamma)/2$.
For autocorrelations with exponent $\alpha(\gamma)\ge 1$
we would no detect anomalous scaling for the
standard deviation ($\beta(\gamma)=0.5$).

Our numerical analysis on the NYSE index shows very sharply
that $\delta(L,\gamma)$ has 
an anomalous power-law behaviour 
in the range from one day to one year ($L=250$).
For example, for $\gamma=1$ we find $\beta(1) \simeq 0.19$,
while for $\gamma=1.5$, $\beta(1.5) \simeq 0.27$
(see fig. 1).
For larger $L$ the statistics becomes insufficient.

The crucial result is that $\beta(\gamma)$
is a not constant function of $\gamma$ in the range 
$-0.5<\gamma<+4$ (see fig. 2), 
showing the presence of different scales.
The interpretation is that different 
$\gamma$ select different typical fluctuation sizes,
any of them being power-law correlated
with a different exponent.

\begin{figure}
\mbox{\psfig{file=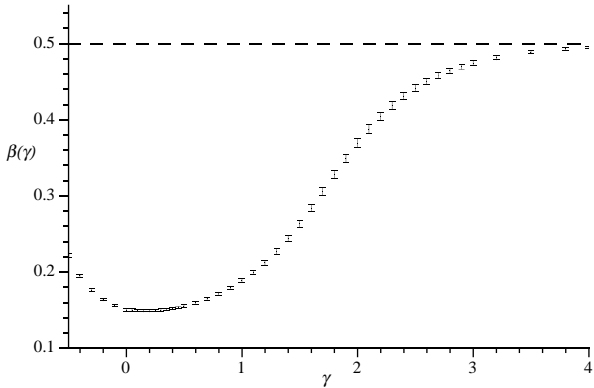,width=8cm,angle=270}}
\caption{Scaling exponent $\beta(\gamma)$ of the standard deviation
$\delta(L,\gamma)$, where the bars represent the errors
over the best fits.
An anomalous scaling $(\beta<0.5)$ is shown 
in the range $-0.5<\gamma<+4$.
}
\end{figure}

The longest correlation is for $\gamma=0.15$ 
($\beta(0.15) \simeq 0.15$).
The case $\gamma=0$ corresponds to
cumulative logarithm of absolute returns.

In the region $\gamma \gtrsim 4$ the averages 
are dominated by only few events, 
corresponding to very large returns, and, therefore,
the statistics becomes insufficient.

The anomalous power-law scaling can be directly tested against 
the plot of autocorrelations.
For instance, the autocorrelations of $r_t$ and of $|r_t|$ 
are plotted in fig. 3 as a function of the correlation length $L$.
Notice that the full line, which is in a good agreement
with the data, is not a plot 
but it is a power-law whose exponent $2 \beta(1)\simeq 0.38$ 
is obtained by the previous scaling analysis of the variance.
The autocorrelations for the return, as expected, vanish 
except for the first step ($L=1$).

It should be also noticed that
a direct analysis of the autocorrelations
would not have provided an analogous
clear evidence for multiscale power-law behaviour, 
since the data show a wide spread 
compatible with different scaling hypothesis.

\begin{figure}
\mbox{\psfig{file=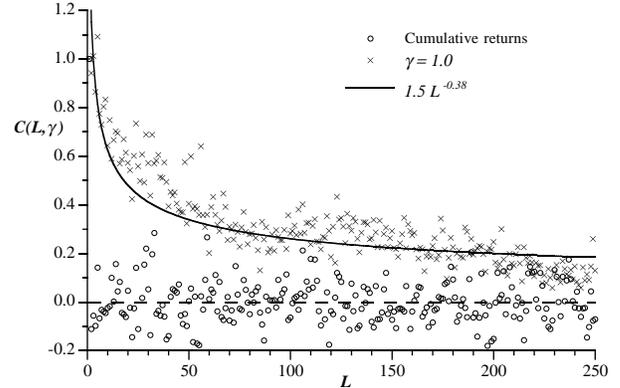,width=8truecm,angle=270}}
\caption{Autocorrelation function of 
$|r_t|$ (crosses) as a function of the correlation length $L$,
 compared with the autocorrelation function of $r_t$ (circles). 
The data are in agreement with a 
power-law with exponent  $2 \beta(1)\simeq 0.38$
in the first case, and absence of correlations in the second. 
In both cases the scale is fixed
by autocorrelations equal to 1 at $L=1$.
}
\end{figure}

In conclusion, we have found that scaling of variance
of the generalized cumulative absolute returns
implies multiscaling
power-law correlations in financial indices.
This result clearly suggests that models 
with exponential correlations, like ARCH-GARCH,
are inadequate to describe
the dynamics of financial markets, and
they should be implemented
to account for the coexistence of
long memory with different scales.

\end{document}